\documentclass[12pt]{iopart}
\usepackage{iopams}
\usepackage[dvips]{graphicx}
\usepackage[dvips]{epsfig,color}
\usepackage{dcolumn}
\usepackage{bm}% bold math

\begin{document}
\title{Role of Hund's coupling in stabilization of the $(0,\pi)$ ordered 
SDW state within the minimal two-band model for iron pnictides}
\author{Nimisha Raghuvanshi and Avinash Singh}
\address{Department of Physics, Indian Institute of Technology Kanpur - 208016}
\ead{avinas@iitk.ac.in}
\begin{abstract}
Spin wave excitations and stability of the $(0,\pi)$ ordered SDW state are investigated within the minimal two-band model for iron pnictides including a Hund's coupling term. The SDW state is shown to be stable in two distinct doping regimes --- finite hole doping in the lower SDW band for small second neighbour hoppings, and small electron doping in the upper SDW band for comparable first and second neighbour hoppings. In both cases, Hund's coupling strongly stabilizes the SDW state due to generation of additional ferromagnetic spin couplings involving the inter-orbital part of the particle-hole propagator. Spin wave energies for the two-band model are very similar to the one-band $t$-$t'$ Hubbard model results obtained earlier, and are in agreement with inelastic neutron scattering studies of iron pnictides. 
\end{abstract}
\maketitle

\section{Introduction}
Single-crystal neutron scattering studies of iron pnictides have indicated a commensurate magnetic ordering of iron moments ordered ferromagnetically in the $b$ direction and antiferromagnetically in the $a$ and $c$ directions \cite{goldman_2008}. Inelastic neutron scattering measurements in $\rm A Fe_2 As_2$ (A = Ca, Ba, Sr) yield well-defined spin-wave excitations up to the zone boundary on an energy scale $\sim 200$meV \cite{zhao_2008,zhao_2009,diallo_2009}. The realization of a $(0,\pi,\pi)$ ordered SDW state has opened the possibility of observing phenomena in this class of compounds which are characteristically associated with both the antiferromagnetic (AF) state such as quantum spin fluctuations, hole/electron motion in AF background, spin-fluctuation mediated pairing as well as the metallic ferromagnetic (F) state such as carrier-induced spin interactions, correlation-induced spin-charge coupling, and non-quasiparticle states. 

%\cite{daghofer_2008,moreo_2009,yu_2009,lu_2009,chen_2009,calderon_2009,kubo_2009, %lee_2009,lv_2010,zhang_2009b}.

A minimal two-band model consisting of two degenerate orbitals $\rm d_{xz}$ and $\rm d_{yz}$ per Fe ion on a two dimensional square lattice \cite{raghu_2008} has been widely studied recently \cite{daghofer_2008,yu_2009,lu_2009,chen_2009,kubo_2009,lv_2010} in order to understand the magnetic ordering and excitations in doped iron pnictides. Ab initio calculations \cite{zhang_2009a} using local density approximation suggest that the Fermi surface is determined by bands having mostly $\rm d_{xz}$ and $\rm d_{yz}$ character, as also indicated by polarized angle-resolved photoemission spectroscopy (ARPES) \cite{zhang_2011}. The hybridization of Fe 3d orbitals with themselves as well as through the As 3p orbitals lying above and below the square plaquettes formed by the Fe atoms leads to effective hopping parameters of the two-orbital model as shown schematically in Fig. 2.
%Illustration of FeAs layer in $\rm Ln O Fe As$ (Ln = La, Ce, Nd, Sm, Gd) and $\rm M Fe_2 As_2$ (A = Ca, Ba, Sr) viewed along the c-axis is given in Fig. 1.

However, the role of inter-orbital exchange interaction (Hund's coupling) on spin wave energies and the stability of the $(0,\pi)$ ordered SDW state have not been investigated within the minimal two-band model. This study would also allow for comparison with the one-band model results obtained earlier within the $t$-$t'$ Hubbard model \cite{faf0,faf1}. The F-AF state in two and three dimensions was shown to be stabilized by the AF spin couplings generated by the second neighbour hopping $t'$ as well as the carrier-induced F spin couplings as in metallic ferromagnets, which are strongly enhanced by the $t'$ induced asymmetric and peaked electronic spectral distribution due to band dispersion saddle points.  

\begin{figure}
\vspace*{-0mm}
\hspace*{0mm}
\psfig{figure=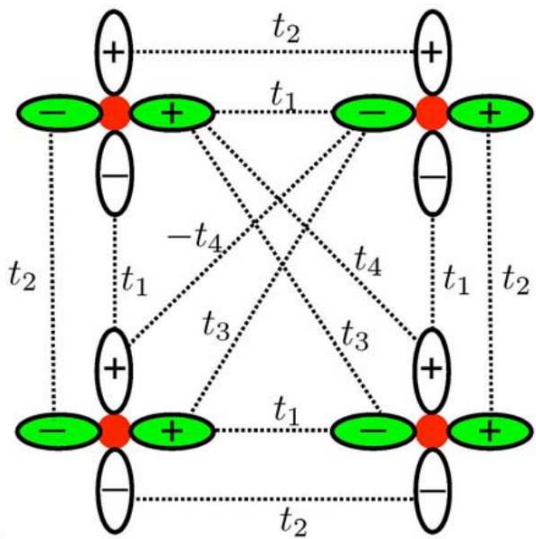,width=75mm,angle=0}
\vspace*{-0mm}
\caption{The effective hopping parameters of the two-band model involving the two orbitals $\rm d_{xz}$ and $\rm d_{yz}$ per Fe ion on a square lattice (from Ref. [5]). The different $\sigma$ and $\pi$ orbital overlaps result in anisotropic intra-orbital hoppings $t_1$ and $t_2$ with opposite sign in different directions. Also included are second neighbour intra-orbital ($t_3$) and inter-orbital ($t_4$) hoppings.}
\end{figure}

\section{$(0,\pi)$ ordered SDW state of the two-band model}
Hund's coupling is known to stabilize metallic ferromagnetism in orbitally degenerate systems \cite{vollhardt_review}. Recent investigations of the correlated motion of electrons in multiband ferromagnets \cite{hunds} have yielded an effective quantum expansion parameter $\frac{[U^2 + ({\cal N}-1)J^2]}{[U+({\cal N}-1)J]^2}$ in terms of the orbital degeneracy ${\cal N}$, intrasite Coulomb interaction $U$, and Hund's coupling $J$, which provides an effective measure of the correlation induced quantum corrections which reduce the spin stiffness and spin wave energies and thus destabilize the ferromagnetic state. The strong suppression of this quantum parameter by Hund's coupling, especially for large ${\cal N}$, provides insight into the stabilization of metallic ferromagnetism in orbitally degenerate systems. As these quantum corrections involving self energy and vertex corrections will affect the ferromagnetic spin couplings, the role of Hund's coupling on magnetic excitations in doped iron pnictides is therefore of interest. 

The inter-orbital Hund's coupling term can be conveniently included on an equal footing with the intra-orbital Hubbard interaction term in a general multi-orbital correlated electron model:
\begin{equation}
\fl
H = -\sum_{\langle ij\rangle \mu\nu\sigma} t_{ij}^{\mu\nu} 
(a_{i\mu\sigma}^\dagger a_{j\nu\sigma} + a_{j\nu\sigma}^\dagger a_{i\mu\sigma}) 
- \sum_{i \mu \nu} U_{\mu \nu} {\bf S}_{i \mu} \cdot {\bf S}_{i \nu}
\end{equation}
where the interaction matrix elements $U_{\mu \nu} = U_{\mu}$ for $\mu = \nu$ and $U_{\mu \nu} = 2J$ for $\mu \ne \nu$ refer to the intra-orbital and inter-orbital Coulomb interaction terms, respectively. In the following, we will specifically consider a two-orbital model on a square lattice with reference to the $\rm d_{xz}$ and $\rm d_{yz}$ orbitals of interest for iron pnictides, with the four hopping terms $t_1$ - $t_4$ as shown in Fig. 1. The inter-orbital density interaction term $Vn_{i\mu} n_{i\nu}$ has been dropped as it does not play any role in the magnetism up to the random phase approximation (RPA) considered here. 

We consider the $(0,\pi)$ ordered SDW state of the above two-band model, with F and AF spin orderings along the $x$ and $y$ directions, respectively. For the single-band model, this SDW state can be conveniently represented in a two-sublattice basis \cite{faf0,faf1}. Extending this approach to a composite two-orbital ($\alpha$  $\beta$), two-sublattice (A  B) basis, the Hartree-Fock (HF) level Hamiltonian matrix in this composite basis (A$\alpha$  A$\beta$  B$\alpha$  B$\beta$) assumes the form:
\begin{equation}
\fl
H_{\rm HF}^{\sigma} ({\bf k}) = \left [ \begin{array}{cccc} -\sigma \Delta_\alpha + \epsilon_{\bf k}^{1x} \;\;
& 0 & \epsilon_{\bf k}^{2y} + \epsilon_{\bf k}^{3} & \epsilon_{\bf k}^{4} \\ 
0 \;\; & -\sigma \Delta_\beta + \epsilon_{\bf k}^{2x} & \epsilon_{\bf k}^{4} & \epsilon_{\bf k}^{1y}+\epsilon_{\bf k}^{3}\\ 
\epsilon_{\bf k}^{2y} +  \epsilon_{\bf k}^{3} \;\; & \epsilon_{\bf k}^{4} &\sigma \Delta_\alpha +\epsilon_{\bf k}^{1x} & 0\\ 
\epsilon_{\bf k}^{4}\;\; & \epsilon_{\bf k}^{1y}  + \epsilon_{\bf k}^{3} & 0 & \sigma \Delta_\beta + \epsilon_{\bf k}^{2x} \end{array} \right ]
\end{equation}
where $\alpha$ and $\beta$ refer to the $\rm d_{xz}$ and $\rm d_{yz}$ orbitals, the band energies:
\begin{eqnarray}
\fl & & \epsilon_{\bf k}^{1x} = -2 t_1 \cos k_x  \;\;\;\;\;\; \epsilon_{\bf k}^{2x} = -2 t_2 \cos k_x  \nonumber \\
\fl & & \epsilon_{\bf k}^{1y} = -2 t_1 \cos k_y  \;\;\;\;\;\; \epsilon_{\bf k}^{2y} = -2 t_2 \cos k_y  \nonumber \\
\fl & & \epsilon_{\bf k}^{3} = -4 t_3 \cos k_x \cos k_y  \;\;\;\;\;\; \epsilon_{\bf k}^{4} = -4 t_4 \sin k_x \sin k_y 
\end{eqnarray}
corresponding to different hopping terms in different directions, and the self-consistently determined exchange fields:
\begin{eqnarray}
\fl
2\Delta_\alpha &=& U_{\alpha} m_{\alpha} + Jm_{\beta} \nonumber \\
\fl 
2\Delta_\beta &=& U_{\beta} m_{\beta} + Jm_{\alpha}
\end{eqnarray}
in terms of the sublattice magnetizations $ m_\alpha$ and $ m_\beta$ for the two orbitals. Quantum corrections to sublattice magnetization resulting from inter-band spectral weight transfer due to electron-magnon interaction should contribute to the substantially reduced magnetic moment observed in iron pnictides.  

%\section{Limiting cases}
It is instructive to consider two limiting cases which connect to the SDW state of the one-band model. When $t_1 = t_2 = t$, $t_3 = t'$, and $t_4 = 0$, the two orbitals in Eq. (2) get decoupled and are also exactly degenerate as $t_1=t_2$, and Eq. (2) identically reduces to the SDW state Hamiltonian for the one-band $t$-$t'$ Hubbard model \cite{faf1}:
\begin{equation}
\fl
H_{\rm HF}^\sigma ({\bf k}) = \left [ \begin{array}{cc} -\sigma \Delta + \epsilon_{\bf k} ^x  \;\;
& \epsilon_{\bf k} ^y +  {\epsilon'_{\bf k}} \\ \epsilon_{\bf k} ^{y}  + {\epsilon'_{\bf k}} \;\;
& \sigma \Delta + \epsilon_{\bf k} ^x  \end{array} \right ]
\end{equation}
where $\epsilon_{\bf k} ^{x(y)} = -2t\cos k_{x(y)} $ and $\epsilon'_{\bf k} = - 4t' \cos k_x \cos k_y$. 

In the more relevant limiting case $t_1 = -t_2 = -t$, $t_3 = -t'$, and $t_4 = 0$, although the two orbitals in Eq. (2) again get decoupled, they are no longer degenerate as $t_1 \ne t_2$. For the $\beta$ orbital, although the off-diagonal terms have exactly opposite sign as compared to Eq. (5) for the one-band model due to the $t_1$, $t_3$ sign reversal, the energies and amplitudes remain unchanged due to energy band folding in the SDW state. For the $\alpha$ orbital, the transformation $k_x \rightarrow  k_x + \pi$ cancels the effect of the $t_1$, $t_3$ sign reversal on both diagonal and off-diagonal terms, and again renders the SDW state identical to Eq. (5) for the one-band model. 

The above equivalence also implies identical spin-wave dispersion as for the one-band model. In the following, we will investigate the effects of finite band mixing term $t_4$ on the spin wave energies in the $(0,\pi)$ ordered SDW state. 

\section{Transverse spin fluctuations in the broken-symmetry state}
In the random phase approximation, the transverse spin fluctuation propagator for the two-band model includes both $U$ and $J$ ladders, and hence retains its usual form:
\begin{equation}
\fl
[\chi^{-+} _{\rm RPA} ({\bf q},\omega)] = \frac{[\chi^0 ({\bf q},\omega)]}{{\bf 1} - [U][\chi^0 ({\bf q},\omega)]}
\end{equation}
where the interaction matrix $[U]$ includes the Hund's coupling term $J$ as off-diagonal matrix elements, as given below (1). Spin wave energies were obtained approximately from poles of Eq. (6) in terms of the largest eigenvalue of real part of $[U][\chi^0]$. The bare particle-hole propagator:
\begin{equation}
\fl
[\chi^0 ({\bf q},\omega)]_{ab} = \sum_{k,l,m} \left [ {\frac{{\phi^{a}_{{\bf k} \uparrow l}}{\phi^{b}_{{\bf k} \uparrow l}}{\phi^{a}_{{\bf {k-q}} \downarrow m}}{\phi^{b}_{{\bf {k-q}} \downarrow m}}}{E^+_{{\bf {k-q}} \downarrow m} - E^-_{{\bf k} \uparrow l} + \omega -i\eta}} + {\frac{{\phi^{a}_{{\bf k} \uparrow l}}{\phi^{b}_{{\bf k} \uparrow l}}{\phi^{a}_{{\bf {k-q}} \downarrow m}}{\phi^{b}_{{\bf {k-q}} \downarrow m}}}{E^+_{{\bf k} \uparrow l} - E^-_{{\bf {k-q}} \downarrow m} - \omega -i\eta}} \right ]
\end{equation}
is evaluated in the orbital-sublattice basis by integrating out the fermions in the $(0, \pi) $ ordered SDW state. 
Here $E_{{\bf k}\sigma}$ and $\phi_{{\bf k}\sigma}$ are the eigenvalues and eigenvectors of the Hamiltonian matrix (2), the orbital-sublattice basis indices $a,b$ run through 1 to 4, and $l,m$ indicate the four eigenvalue branches. The superscripts $+ (-)$ refer to particle (hole) energies above (below) the Fermi energy, and both inter-band and intra-band particle-hole terms are included. 

\begin{figure}
\vspace*{-0mm}
\hspace*{0mm}
\psfig{figure=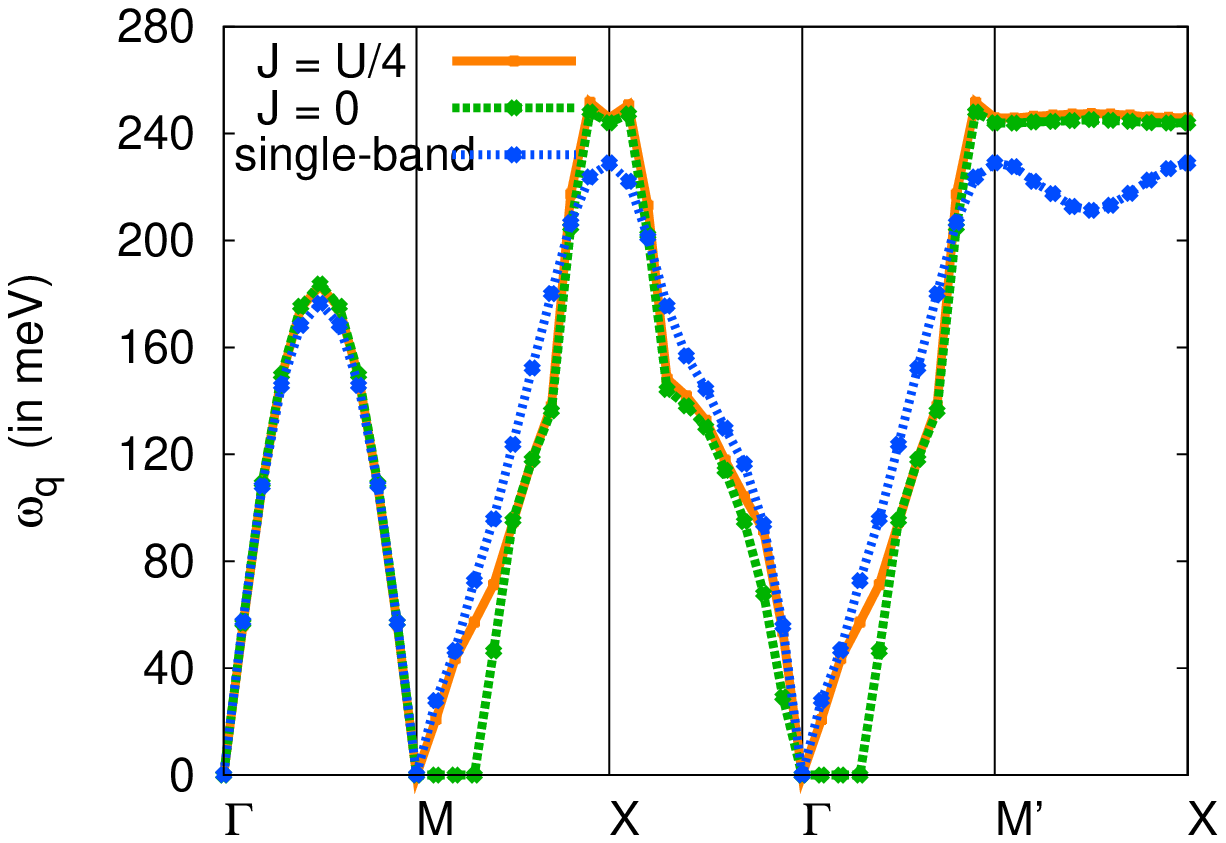,width=100mm,angle=0}
\hspace{-15mm}
\psfig{figure=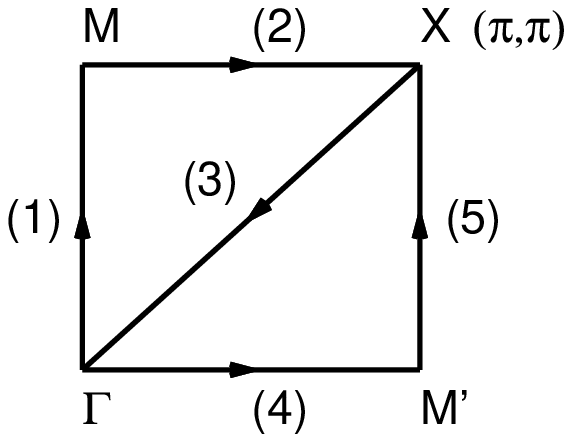,width=60mm,angle=0}
\vspace*{-0mm}
\caption{The spin-wave dispersion obtained for the two-band model with Hund's coupling ($J=U/4$), without Hund's coupling ($J=0$), and for the single-band case ($t_4 = 0$). Here $t_3$=$t_4$=$-0.3$, and the hopping energy scale $|t_1| = 200$meV.}
\end{figure}

The bare propagator $[\chi^0]$ has a finite imaginary part representing (low-energy) intra-band and (high-energy) inter-band particle-hole excitations, resulting in finite spin wave damping and linewidth even at the RPA level, as in metallic antiferromagnets \cite{met_af}. The spin-charge coupling mechanism relevant for metallic ferromagnets such as manganites will also contribute due to decay of spin waves into longer wavelength modes accompanied with internal charge excitations \cite{spchg}. Indeed, this has been suggested from the absence of any steep increase in damping at higher energy indicative of a Stoner continuum \cite{zhao_2009}. From the observed high energy behaviour of spin wave damping ascribed to particle-hole excitations \cite{diallo_2009}, it has been inferred that the full excitation spectrum can not be understood in terms of the local moment picture. Weakly damped spin waves near the ordering wavevector have been obtained within multiband models from the imaginary part of the spin fluctuation propagator \cite{brydon_2009,knolle_2010,cvetkovic_2009}.

Fig. 2 shows the spin-wave dispersion in the $(0,\pi)$ state of the two-band model, showing the stabilization of the F-AF state for small Hund's coupling. For $J=0$, the spin wave energy becomes negative for small $q_x$ near the M and $\Gamma$ points, indicating instability with respect to long wavelength spin twisting modes in the ferromagnetic ordering direction. Here, the hopping terms are $t_1$=$-1.0$, $t_2$=1.0, and $t_3$=$t_4$=$-0.3$, the exchange field $\Delta$=3.0, and the hole doping concentration $x\sim 35\%$, for which the self-consistency condition (4) yields  $U+J \sim 11$. Also shown is the single-band result obtained by simply setting $t_4=0$. We have set $|t_1|=1$ as the unit of energy scale, and the spin wave energies are shown for $|t_1|$=200meV. 

Evidently, the electronic spectral function modification due to the band mixing term $t_4$ reduces the F spin couplings and destabilizes the F-AF state. The additional spin couplings $J^2 [\chi^0]_{ij} ^{\alpha\beta}$ generated by Hund's coupling due to the inter-orbital component of the particle-hole propagator restores the stability of the F-AF state. 

Fig. 2 also shows that for small $t_4$, the overall structure of the spin wave dispersion for the two-band model, with  Hund's coupling included, is very close to that for the one-band model obtained earlier,\cite{faf1} with respect to anisotropic spin wave velocities, spin-wave dispersion, and energy scale. Significantly, the spin-wave dispersion clearly shows a maximum at $(\pi,\pi)$ in agreement with neutron scattering experiments \cite{zhao_2008,zhao_2009,diallo_2009}.

\begin{figure}
\vspace*{-0mm}
\hspace*{0mm}
\psfig{figure=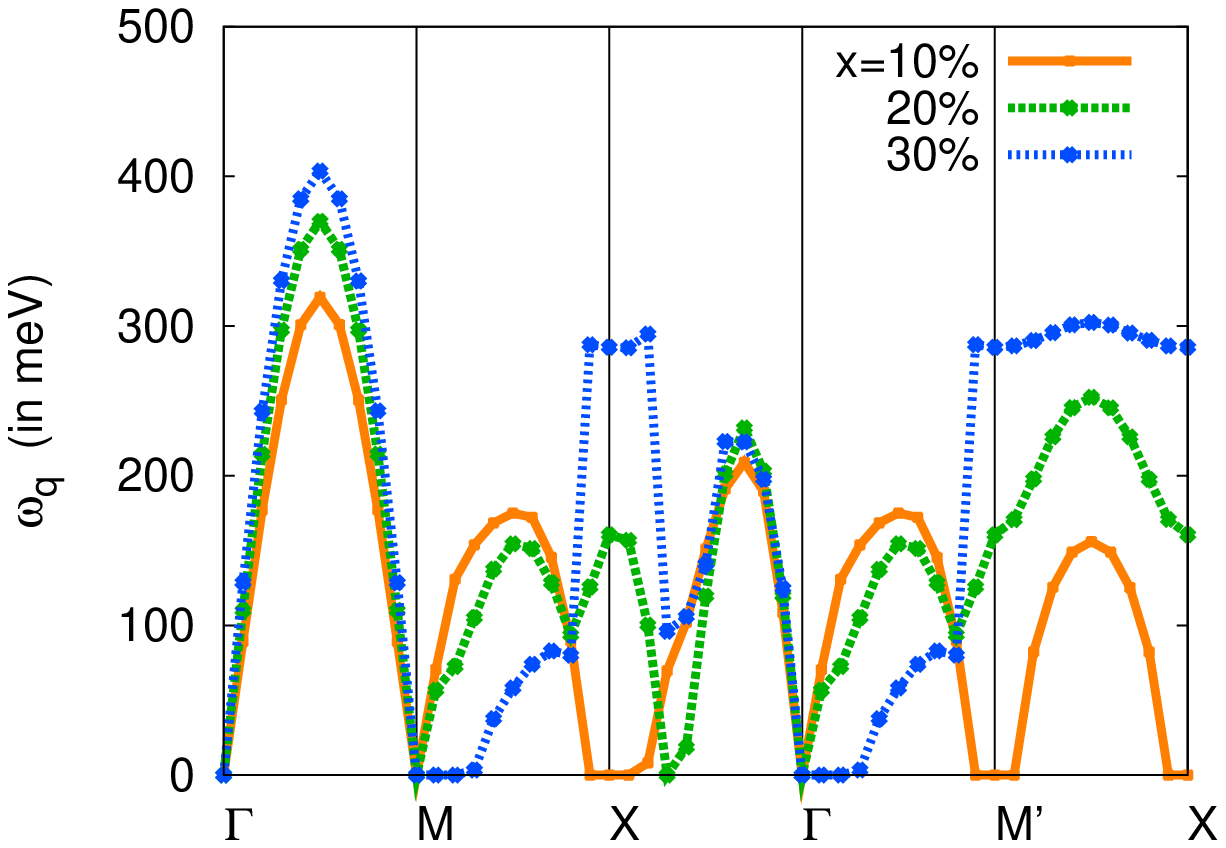,width=80mm,angle=0}
\psfig{figure=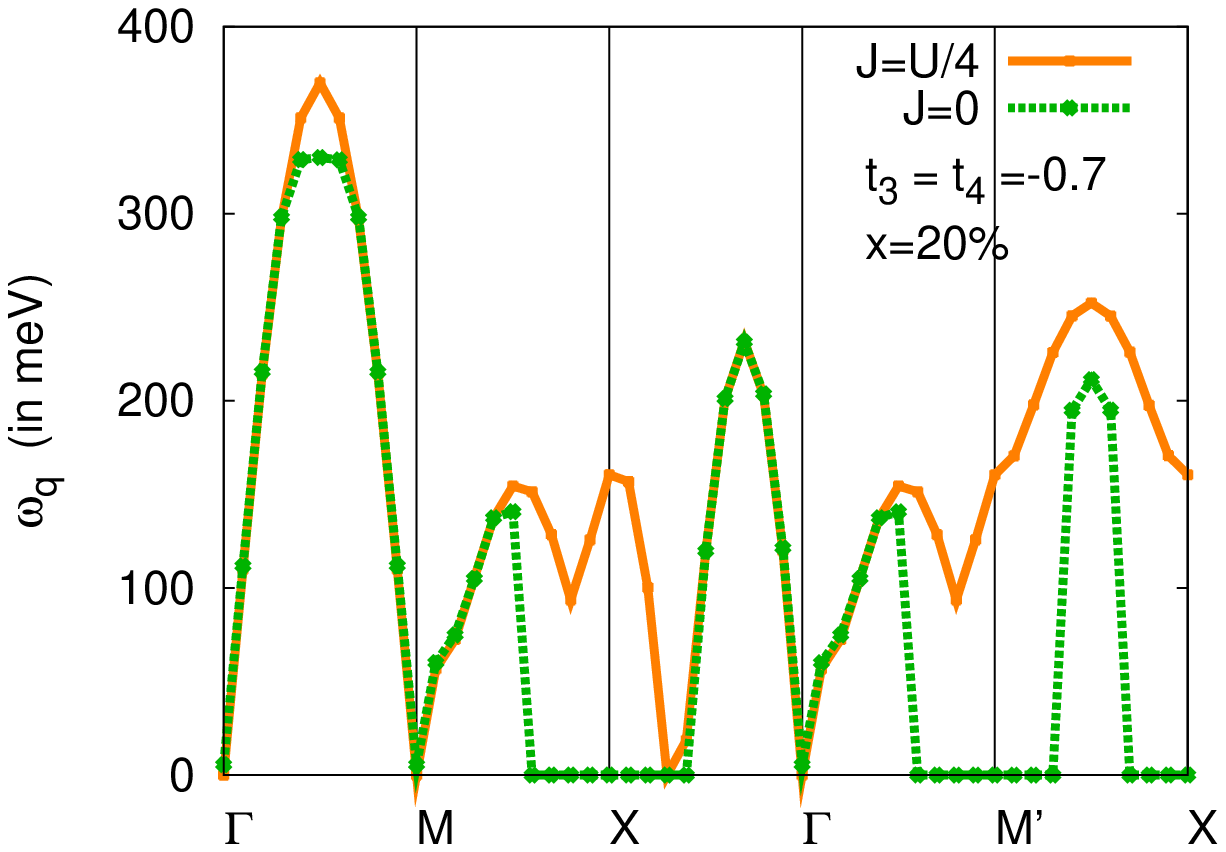,width=80mm,angle=0}
\vspace*{-0mm}
\caption{(a) Hole doping dependence of spin-wave dispersion for the two-band model with Hund's coupling included. Here $t_3$=$t_4$=$-0.7$ and $\Delta = 2$. (b) Without Hund's coupling ($J=0$), the spin wave energies become negative over a larger part of the Brillouin zone.}
\end{figure}

For larger second-neighbour hoppings $t_3$=$t_4$=$-0.7$ and somewhat smaller exchange field $\Delta$=2, the hole doping  dependence of spin wave dispersion is shown in Fig. 3(a). The self-consistency condition yields $U+J \sim 8$ for $x=20\%$. While the peak spin wave energy in the AF direction is indeed enhanced, as expected from the stronger second-neighbour AF spin couplings generated, the F-AF state is evidently not robust with respect to fluctuations in the F ordering direction. For $x=10\%$, the spin wave energy becomes negative near X and M', indicating instability with respect to zone boundary modes, whereas at higher hole doping $x=30\%$, the spin wave energy becomes negative near $\Gamma$ and M, indicating instability with respect to long wavelength modes. 

For the relatively stable case at $x=20\%$, the spin wave energies with and without Hund's coupling are compared in Fig. 3(b). Again, Hund's coupling is seen to play a crucial role in stabilizing the F-AF state, as indicated by the negative spin wave energies obtained for $J$=0 over a larger part of the Brillouin zone.  
%Overall, while the AF spin couplings are enhanced at higher $t_3$, $t_4$, the F spin couplings get significantly weakened. 

\begin{figure}
\vspace*{-0mm}
\hspace*{0mm}
\psfig{figure=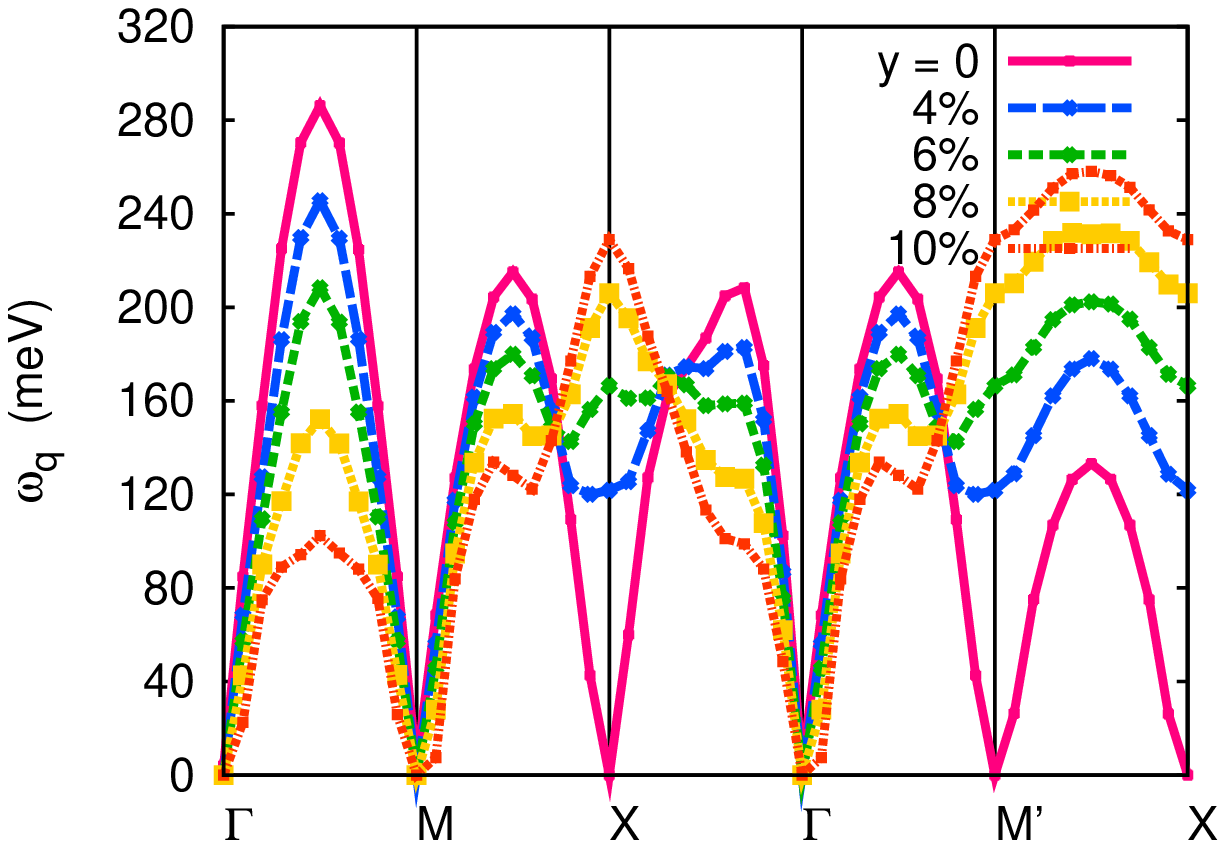,width=80mm,angle=0}
\psfig{figure=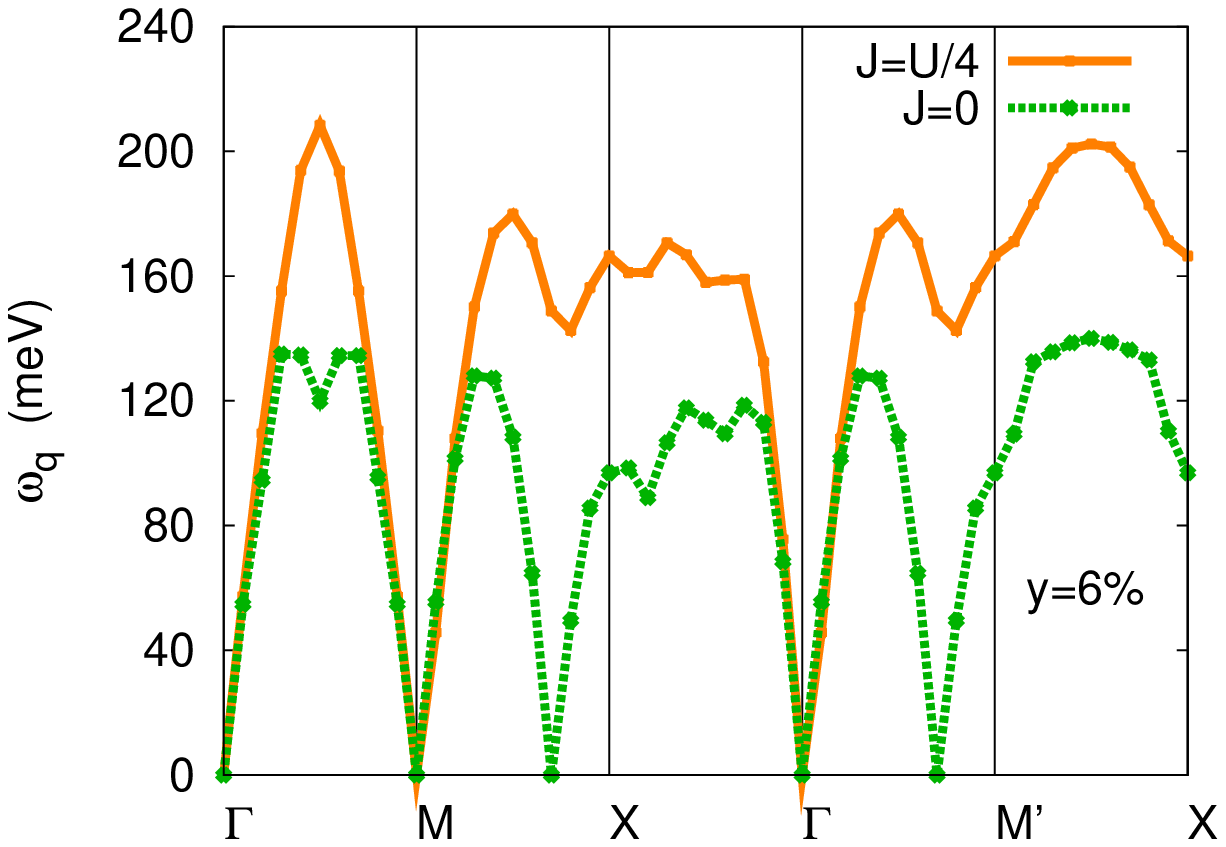,width=80mm,angle=0}
\vspace*{-0mm}
\caption{(a) Spin wave energies for the two-band model with same hopping parameters as in Ref. [5], showing strong enhancement (suppression) in the F (AF) ordering direction with increasing electron doping $y$. (b) Spin wave energies are strongly enhanced when a small Hund's coupling term is included.}
\end{figure}

Finally, we consider the set of hopping parameters $t_1$=$-1.0$, $t_2$=1.3, $t_3$=$t_4$=$-0.85$, which yields circular electron and hole pockets near the bottom of the upper band.\cite{raghu_2008} At half filling, the spin wave dispersion 
(Fig. 4(a), $y=0$) is very similar to that derived for the single band $t$-$t'$ model \cite{faf0}, for which the spin wave energy vanishes at the zone boundary in the ferromagnetic direction due to absence of any ferromagnetic spin coupling in the insulating state. 

For small electron doping in the upper SDW band, finite F spin couplings are generated which further stabilize the F-AF state, as seen from the spin wave dispersion in Fig. 4(a). Here $\Delta$=3 and a small Hund's coupling has been included. At $y=8\%$, the spin wave energy at the F zone boundary X exceeds that at the AF zone boundary, as observed experimentally. The F-AF state is seen to be stable for very low electron doping, and gets rapidly destabilized with increasing electron doping due to suppression of AF spin couplings. This behaviour is in agreement with the observed rapid suppression of magnetic order in iron pnictides with electron doping (due to F substitution of O atoms in $\rm LaO_{1-x}F_x FeAs$ or Ni substitution of Fe atoms in $\rm BaFe_{2-x} Ni_x As_2$) \cite{wang_2010}.

Notably, the spin wave dispersion, energy scale, and doping behaviour for the electron doped SDW state are all very similar to Fig. 2 for the hole doped SDW state as well as the single-band model. Evidently, it is the generation of the ferromagnetic spin couplings in all three cases which is the common crucial factor in stabilizing the $(0,\pi)$ ordered SDW state. 
For $\Delta$=2, the SDW state effective gap $\sim 400$meV is well above the maximum spin wave energy, which prevents spin wave excitations from rapidly decaying into a particle-hole continuum, as indeed not observed experimentally up to energies of 200 meV.

\section{Conclusions}
Spin wave energies were obtained in the $(0,\pi)$ ordered SDW state of the minimal two-band model for iron pnictides including the Hund's coupling term between the degenerate $\rm d_{xz}$ and $\rm d_{yz}$ Fe orbitals. Negative spin wave energies were taken as signalling instability of the $(0,\pi)$ SDW state with respect to transverse spin fluctuations, indicating significantly weakened F or AF spin couplings. A robust $(0,\pi)$ SDW state was found in two distinct doping regimes --- finite hole doping in the lower SDW band for small second neighbour hoppings (squarish electron/hole pockets), and small electron doping in the upper band for comparable hopping terms (circular electron/hole pockets). 

In both cases, the Hund's coupling term $J$ was found to strongly stabilize the $(0,\pi)$ SDW state due to the generation of additional ferromagnetic spin couplings $J^2 [\chi^0]_{ij} ^{\alpha\beta}$ involving the inter-orbital part of the particle-hole propagator. Furthermore, the spin wave dispersion, energy scale, and doping behaviour in both cases were found to be very similar to that for the one-band model, which was ascribed to the carrier-induced ferromagnetic spin couplings as the common crucial factor in stabilizing the $(0,\pi)$ ordered SDW state. The emergence of F spin couplings at finite hole/electron doping results in a characteristic peak spin-wave energy at $(\pi,\pi)$, in agreement with neutron scattering experiments on iron pnictides, and also accounts for the large planar anisotropy observed between the AF and F spin couplings in the ab plane. 
 
Evidence of Fermi surface folding associated with the SDW state has been observed in recent ARPES studies \cite{folding_arpes}. Electronic quasiparticle dispersion and spectral function renormalization in the $(0,\pi)$ ordered SDW state due to electron-magnon interaction and multiple magnon emission-absorption processes should therefore be of interest.

\end{document}